# A collision in 2009 as the origin of the debris trail of asteroid P/2010 A2

Colin Snodgrass[1,2], Cecilia Tubiana[1], Jean-Baptiste Vincent[1], Holger Sierks[1], Stubbe Hviid[1], Richard Moissl[1], Hermann Boehnhardt[1], Cesare Barbieri[3], Detlef Koschny[4], Philippe Lamy[5], Hans Rickman[6,7], Rafael Rodrigo[8], Benoît Carry[9], Stephen C. Lowry[10], Ryan J. M. Laird[10], Paul R. Weissman[11], Alan Fitzsimmons[12], Simone Marchi[3] and the OSIRIS team[*]

*[1]Max-Planck-Institut fuer Sonnensystemforschung, Max-Planck-Str. 2, 37191 Katlenburg-Lindau, Germany, [2]European Southern Observatory, Alonso de Córdova 3107, Casilla 19001, Santiago 19, Chile, [3]University of Padova, Department of Astronomy, Vicolo dell'Osservatorio 3, 35122 Padova, Italy, [4]Research and Scientific Support Department, European Space Agency, Keplerlaan 1, Postbus 229, 2201 AZ Noordwijk ZH, Netherlands, [5]Laboratoire d'Astrophysique de Marseille, UMR6110 CNRS/Université Aix-Marseille, 38 rue Frédéric Joliot-Curie, 13388 Marseille Cedex 13, France, [6]Department of Astronomy and Space Physics, Uppsala University, Box 516, 75120 Uppsala, Sweden, [7]PAS Space Research Center, Bartycka 18A, 00-716 Warszawa, Poland, [8]Instituto de Astrofísica de Andalucía, CSIC, Box 3004, 18080 Granada, Spain, [9]LESIA, Observatoire de Paris-Meudon, 5 place Jules Janssen, 92195 Meudon Cedex, France, [10]Centre for Astrophysics and Planetary Science, University of Kent, Canterbury CT2 7NH, UK, [11]Jet Propulsion Laboratory, 4800 Oak Grove Drive, MS 183-301, Pasadena, CA 91101, USA, [12]Astrophysics Research Centre, Queen's University Belfast, BT7 1NN, UK,*

---

[*] Lists of participants and affiliations appear at the end of the paper.

**The peculiar object P/2010 A2 was discovered by the LINEAR near-Earth asteroid survey in January 2010[1] and given a cometary designation due to the presence of a trail of material, although there was no central condensation or coma. The appearance of this object, in an asteroidal orbit (small eccentricity and inclination) in the inner main asteroid belt attracted attention as a potential new member of the recently recognized class of 'Main Belt Comets' (MBCs)[2]. If confirmed, this new object would greatly expand the range in heliocentric distance over which MBCs are found. Here we present observations taken from the unique viewing geometry provided by ESA's Rosetta spacecraft, far from the Earth, that demonstrate that the trail is due to a single event rather than a period of cometary activity, in agreement with independent results from the Hubble Space Telescope (HST)[3]. The trail is made up of relatively large particles of millimetre to centimetre size that remain close to the parent asteroid. The shape of the trail can be explained by an initial impact ejecting large clumps of debris that disintegrated and dispersed almost immediately. We determine that this was an asteroid collision that occurred around February 10, 2009.**

P/2010 A2 orbits much closer to the Sun (semi-major axis = 2.29 AU) than the previously discovered MBCs, whose activity seems to be driven by episodic ice sublimation[2]. The discovery of a parent body a few arc-seconds (~1500 km) away from the trail[4,5] implied that it was debris from a recent collision rather than the tail of a comet, although Earth based observations alone are consistent with a comet model[6]. It was suggested that the trail formed between January and August 2009, and was comprised of relatively large (diameter > 1 mm) grains[7]. Here we use the term "trail" to describe a tail made up of large particles, rather than dust from a currently active comet. HST observations refine the diameter of the parent body to 120 m and the date to February/March 2009[3].

We obtained an improved 3-D description of the trail geometry by observing it with the OSIRIS Narrow Angle Camera[8] on board ESA's Rosetta spacecraft on March 16, 2010. Rosetta was approaching the asteroid belt for its July 2010 fly-by of asteroid 21 Lutetia, and at the time of observation was 1.8 AU from the Sun and 10° out of P/2010 A2's orbital plane. From this vantage point the separation between the anti-velocity (orbit) angle and the anti-Sun (comet tail) direction was much larger than was possible to observe from Earth. We also obtained reference images of P/2010 A2 from Earth using the 3.6 m New Technology Telescope (NTT) at ESO's La Silla observatory and the 200" Hale telescope at Palomar Mountain. Figure 1 displays images of P/2010 A2 at four epochs, from the Earth and from Rosetta. We measured the position angle (PA) of the trail and extracted the flux profile along the trail axis at each epoch (Fig. 2).

We simulate the shape of the observed trail at each epoch by modelling the trajectories of dust grains, as is commonly done for comet tails[9,10]. The motion depends on the grains' initial velocity and the ratio $\beta$ between solar radiation pressure and solar gravity, which is related to the size of the grains[11]. Due to the small phase angle as viewed from Earth it is not possible to find a unique solution for the dust ejection epochs from the ground-based observations alone: The best estimate indicates that particles must have been emitted before August 2009, and should be of at least millimetre size to account for the low dispersion and their apparent position close to the projected anti-velocity vector. The higher phase angle of the OSIRIS observations allows a more precise simulation of the trail, and consequently we obtained a very narrow time frame for the emission of the dust. The grains must have been released around 10 February 2009, plus or minus 5 days, with the uncertainty being due to the measurement of the PA of the faint trail in the OSIRIS images. In order to account for the PA and the length of the trail, we must consider grains ranging from millimetre to centimetre size and larger. The particle sizes from this model together with the brightness profile shown in Fig. 2 allow us to measure the size distribution of grains, and from this derive a total mass of the ejecta of 3.7 x $10^8$ kg, or approximately 16% of a 120 m diameter parent body, assuming a density of 2500 kg m$^{-3}$ and an albedo of 15% for both the asteroid and the grains.

The shape of the trail cannot be reproduced with a traditional comet tail model, even when considering a longer time scale for the event. Cometary models all produce tail geometries in the OSIRIS image with a fan that reaches a point at the nucleus and becomes wider farther from it (see supplementary material for examples). All images of P/2010 A2 show a distinctive broad edge at the 'nucleus' end and then a trail with parallel edges. From the Rosetta observing geometry this edge is even broader than it is from Earth. This shape can be reproduced by a number of parallel synchrones, representing dust produced at the same time. In this model, an initial dust cloud is formed (presumably by a collision) in February 2009, which initially does not spread much (less than 1000 km) but over a year solar gravity and radiation pressure expand this small trail to its observed width and length, respectively. Higher resolution images from HST[3] show the presence of parallel striae in the trail, very well aligned with the synchrone representing the original event as estimated from our simulations. These striae indicate that some areas of higher densities existed in the original cloud; larger clumps of material which fragmented and dispersed as they were ejected. The width of the broad front end of the trail from these different geometries can be used to constrain the speed of particles in the original ejecta cloud to less than 1 m s$^{-1}$. Impact experiments[12] find that such a low velocity implies a low strength and high porosity parent body, although recent computer simulations suggest that impacts on such a small asteroid will lead to low velocity ejecta independent of porosity[13].

Previously, asteroid collision models have been used to explain the dust trails associated with MBCs[14], but the longer lasting dust production and repeated activity of comet Elst-Pizarro at each perihelion[15,16] rule out recent collisions (where 'recent' means within the past few years). Collisions inferred from asteroid families[17] or large scale denser regions in the zodiacal dust cloud[18] have ages of $10^4$ to $10^9$ years. Our observations show the first direct evidence for a collision that is recent in observational terms, with a debris trail that is still evolving. From estimates of the population of the main asteroid belt[19,20] and an estimated impactor diameter of 6-9 m[(21)], we expect roughly one impact of this size every 1.1 Gyr for a 120 m diameter parent body, or approximately one every 12 years somewhere in the asteroid belt. This is in agreement with a single detection by the LINEAR survey; we expect that more small collisions will be detected by next-generation surveys. Collisions of this size therefore contribute around $3 \times 10^7$ kg yr$^{-1}$ of dust to the zodiacal cloud, which is negligible compared with comets and the total required to maintain a steady state[22], in agreement with recent models[23].

**Supplementary Information** accompanies the paper on **www.nature.com/nature**.

Acknowledgements: We thank Rita Schulz and the Rosetta operations team for enabling these 'target of opportunity' observations to be performed. OSIRIS is funded by the national space agencies ASI, CNES, DLR, the Spanish Space Program (Ministerio de Educacion y Ciencia), SNSB and ESA. The ground-based observations were collected (in part) at the European Southern Observatory, Chile, under programmes 084.C-0594(A) and 185.C-1033(A).

Author Contributions: CS and CT lead this project and performed the data reduction and analysis, JBV did the modelling and lead the interpretation, HS, SH and RM were responsible for the planning and execution of the OSIRIS observations, HB contributed to the modelling and interpretation. CB, DK, PL, HR and RR are the Lead Scientists of the OSIRIS project. The OSIRIS team built and run this instrument and made the observations possible. BC, SL, RL, PW and AF were the observers who provided the ground based observations. SM provided calculations of the collision probability.

Author information: The authors declare no competing financial interests. Correspondence and requests for materials should be addressed to CS (snodgrass@mps.mpg.de).

**The OSIRIS team** M. A'Hearn[13], F. Angrilli[14], A. Barucci[9], J.-L. Bertaux[15], G. Cremonese[16], V. Da Deppo[17], B. Davidsson[6], S. Debei[14], M. De Cecco[18], S. Fornasier[9], P. Gutiérrez[8], W.-H. Ip[19], H. U. Keller[20], J. Knollenberg[21], J R Kramm[1], E. Kuehrt[21], M. Kueppers[22], L. M. Lara[8], M. Lazzarin[3], J. J. López-Moreno[8], F. Marzari[23], H. Michalik[20], G. Naletto[24], L. Sabau[25], N. Thomas[26], K.-P. Wenzel[4]

Affiliations for participants: [13]University of Maryland, Department of Astronomy, College Park, Maryland 20742-2421, USA. [14]Department of Mechanical Engineering - University of Padova, Via Venezia 1, 35131 Padova, Italy. [15]LATMOS, CNRS/UVSQ/IPSL, 11 Boulevard d'Alembert, 78280 Guyancourt, France. [16]INAF - Osservatorio Astronomico di Padova, Vicolo dell'Osservatorio 5, 35122 Padova, Italy. [17]CNR-IFN UOS Padova LUXOR, Via Trasea 7, 35131 Padova, Italy. [18]UNITN, Università di Trento, Via Mesiano, 77, 38100 Trento, Italy. [19]National Central University, Institute of Astronomy, 32054 Chung-Li, Taiwan. [20]Institut für Datentechnik und Kommunikationsnetze der TU Braunschweig, Hans-Sommer-Str. 66, 38106 Braunschweig, Germany. [21]DLR Institute for Planetary Research, Rutherfordstr. 2, 12489 Berlin, Germany. [22]ESA-ESAC, Camino bajo del Castillo S/N, 28691 Villanueva de la Cañada, Madrid, Spain. [23]Department of Physics - University of Padova, Via Marzolo 8, 35131 Padova, Italy. [24]Department of Information Engineering - University of Padova, Via Gradenigo, 6/B I, 35131 Padova, Italy. [25]Instituto Nacional de Tecnica Aeroespacial, Carretera de Ajalvir, p.k. 4, 28850 Torrejon de Ardoz (Madrid), Spain. [26]Physikalisches Institut, Abteilung Weltraumforschung und Planetologie, Universität Bern, Sidlerstr. 5, 3012 Bern, Switzerland.

Figure 1. Images of P/2010 A2 at four epochs. These are, from top to bottom, from the NTT (February), Rosetta (March), Palomar and the NTT (both April), respectively. The scale bars in the lower right of panels a-d show a projected distance of 5 x 10$^4$ km. When possible, we median combined images centred on the object to increase the S/N ratio (relative to a single exposure) of the trail and remove background stars. To isolate the faint dust trail in the OSIRIS data we first subtract an image of the background star field from each frame before shifting the frame based on the motion of the object and then median combining. On the right we show the images overlaid with synchrones generated from the Finson-Probstein model. Numbers indicate estimates of the particle size distribution along the synchrones, derived from the model. The orientation of the images is North up, East left. The compass in the top left of panels e-h shows the direction of the heliocentric velocity vector (orbit) V and the direction to the Sun. The advantage of the Rosetta observing geometry is clear, with the broad head of the trail and obvious difference between the observed PA and the anti-velocity vector apparent in the OSIRIS image. Models based on a period of cometary activity (rather than a single event) or smaller particle sizes produce a significantly different pattern of synchrones in panel f (see supplementary Figures 2-4), that do not fit the observations. The same models all produce similar synchrones to the impact model for panels e, g and h, and therefore cannot be ruled out based on Earth-based data alone.

Figure 2. Flux profiles along the trail. The normalised profiles for the February NTT (solid black line) and the OSIRIS datasets (dot-dashed red line) are shown. The x-axis is in km along the trail, with the conversion from the projected scale in arc-seconds on sky based on the geometry derived from our model. The vertical dashed lines indicate the Half Maximum (HM) of the profiles, used to measure the scale length of the trails in these images with different sensitivities.

The two profiles have scale lengths of 4.3 x $10^4$ and 9.3 x $10^4$ km along the trail. The right y-axis shows the calibrated surface brightness of the NTT profile in R-band magnitudes per square arc-second. The flux profiles from the other Earth based observations match the NTT one, but are omitted for clarity as they have higher noise due to the shorter integration times. We derive a size distribution using the NTT flux profile and the size of particles as a function of distance along the trail from the Finson-Probstein model. This is done by converting the total flux across the trail at each distance to a reflecting area (assuming an albedo of 15%), and finding the corresponding number of particles of the appropriate size. The resulting cumulative size distribution is shown in supplementary Fig. 6, and has a slope that matches the prediction for a population of collisional remnants[24].

## IMAGE

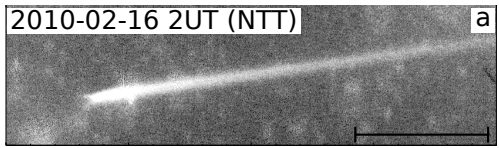

2010-02-16 2UT (NTT) **a**

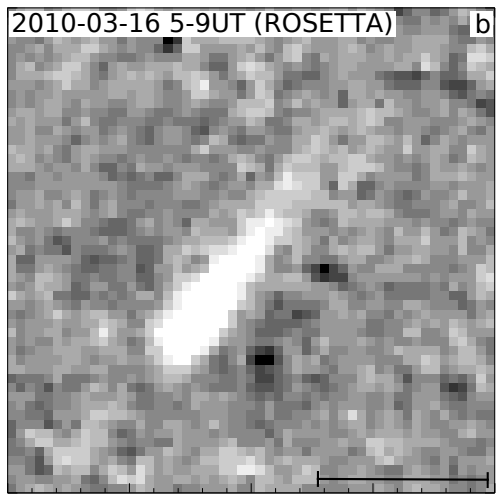

2010-03-16 5-9UT (ROSETTA) **b**

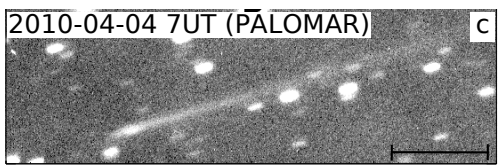

2010-04-04 7UT (PALOMAR) **c**

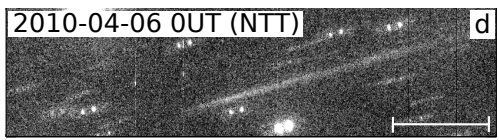

2010-04-06 0UT (NTT) **d**

## IMAGE + MODEL

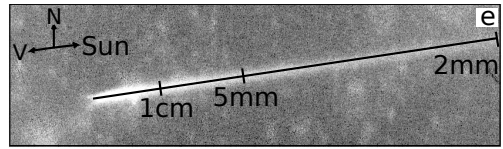

N
V ← Sun
1cm   5mm   2mm   **e**

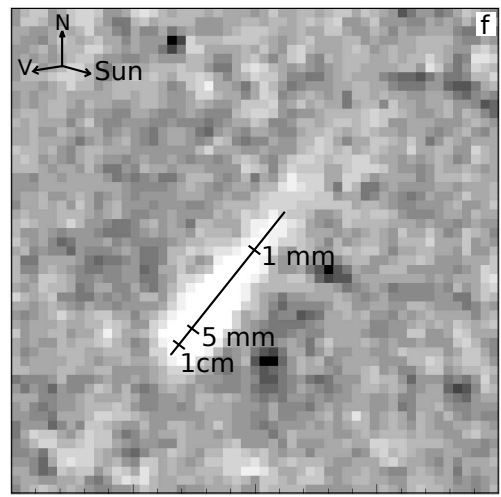

N
V ← Sun
1 mm
5 mm
1cm   **f**

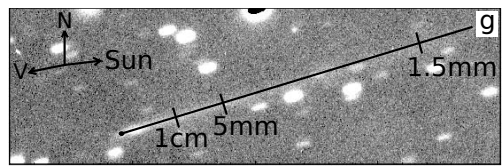

N
V ← Sun
1.5mm
1cm   5mm   **g**

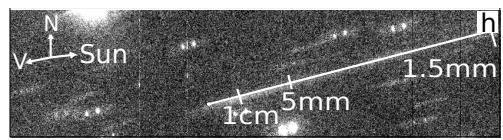

N
V → Sun
1.5mm
1cm   5mm   **h**

Arcsec

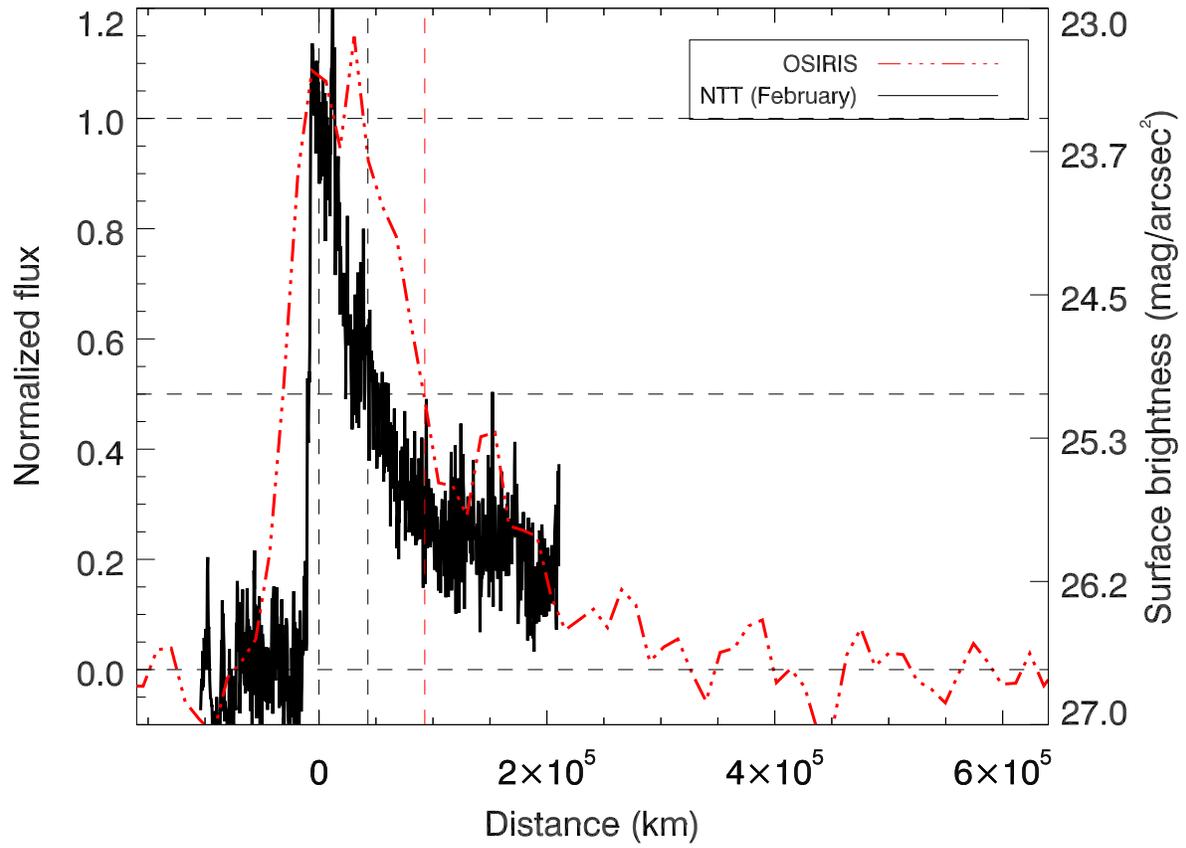

**Supplementary material**

**Observation details** The geometry of observation at each epoch is described in Supplementary Table 1 and illustrated in Supplementary Figure 1. It is clear that from the Earth the viewing geometry remains similar throughout the period of observations, while Rosetta gave a significantly different phase angle and orbital plane angle. All observations (space- and ground-based) were performed with the telescope tracking at the apparent rate of motion of the object. Both the ground based telescopes and Rosetta have sufficient tracking accuracy that there was no need to perform any differential guiding; the star trails in individual images show smooth motion with the expected length and direction and therefore the trail is not affected by any artefacts from tracking errors. All data were reduced in the standard way (bias subtraction, flat fielding etc) using IRAF and IDL. The OSIRIS data was further processed using the following steps: 1. Alignment of all frames on the star background. 2. Median combination to produce a high S/N image of the background star field without cosmic rays or moving objects. 3. Subtraction of this background frame from each individual frame. 4. Shifting of individual background subtracted frames based on the rate of motion of P/2010 A2 to align them on the object. 5. Median combination of the shifted frames to remove cosmic rays and leave only P/2010 A2. This technique is often applied to faint comets, but was particularly effective in this case since the point-spread function (PSF) of OSIRIS is not affected by the Earth's atmosphere and hence remains stable. It was not possible to apply this to the ground based data sets presented in this paper since they were taken over short timeframes during which P/2010 A2 did not move sufficiently far against the stellar background to perform step 2.

**Finson-Probstein models** We simulated the shape of the observed trail using the technique of Finson and Probstein that is commonly applied to comet tails; modelling of the trajectories of grains released from the main body[9,10]. Whether the initial release is due to sublimation (cometary activity) or impact will affect the trajectories at distances close to the parent object. However at a distance of more than several object radii the motion of the grains is dominated by solar gravity ($F_{grav}$) and radiation pressure ($F_{rad}$). Both forces vary with the square of the heliocentric distance and act in opposite directions. Therefore the trajectories of the dust grains can be calculated by solving Newton's two-body problem, multiplying the gravity constant by $1 - \beta$ in the equation of motion, where $\beta = F_{rad}/F_{grav}$. The calculated positions of dust grains in the trail with respect to the parent body are then plotted as a grid of so-called synchrones and syndynes projected onto the image plane. Syndynes give the loci of dust particles with the same $\beta$ ratio but emitted at different times; synchrones describe the loci of dust particles emitted at the same time but with different $\beta$. For grains of diameter $d$ larger than 0.1 microns, $\beta$ can be written as a function of the grain size: $\beta = k/d$ where $k$ is a constant for a given material[11].

We show the output of Finson-Probstein modelling of the dust trail for various scenarios in Supplementary Figures 2-4, which demonstrate the need for a very short duration of activity (i.e. a collision) and large particles. None of these can match the observed geometry in the OSIRIS image, leaving only the short duration (impact) and large particle model described in the main paper. Note that it is impossible to tell the difference between these models from the Earth observing geometry. Furthermore, we show in Supplementary Fig. 5 the different synchrones produced for emission at different times around the derived impact date, which demonstrate the different trail position angles that would have been measured in each case, and therefore show the accuracy of our collision date determination.

**Size distribution of ejecta** We generate a size distribution for the ejecta using the $\beta$ values from the Finson-Probstein modelling. To convert these to sizes we assume a constant $k = 4 \times 10^{-7}$, appropriate for silicate (rocky) material which is a reasonable assumption for asteroidal dust. This gives a relationship between the length $l$ along the trail in km (which is found from the projected distance in arc-seconds and the 3D direction of the trail derived from the model) and the particle diameter: $d = 376/l$, for $d$ in metres. This obviously cannot be extrapolated back to very small distances (where the implied particle diameter would be larger than the parent body), but since the pixel scale in the NTT (February) image corresponds to 312 km along the trail we do not resolve this region and thus avoid the problem. We use this size-distance relationship to find the particle size for each pixel along the NTT flux profile shown in Fig. 2. We then find the number of particles by comparing the total reflecting area, given by the flux integrated across the trail and assuming an albedo of 15%, to the area of a single particle of the appropriate size, which gives the number of particles as a function of particle size. We plot the cumulative size distribution (CSD) in Supplementary Fig. 6, using the usual convention of plotting the number of particles $N(> r)$ larger than a given radius $r$ against the radius, on logarithmic scales. On this log-log plot the power law describing $N(> r)$ as a function of $r^{-q}$ produces a straight line; we find $q = 2.5$ matching the theoretical slope for a population of collisional fragments[24]. We note that the uncertainty on the width of the trail ($17 \pm 1$ pixels in the NTT image) introduces only a small uncertainty in the size distribution. The uncertainty in the conversion from $\beta$ to particle size, where we have to make assumptions about the material, is also small. The difference in the particle size at a given length along the trail is, for extreme cases, a factor of two. Particles are larger at a given distance for a very light material such as graphite that is affected more by Solar radiation pressure than by gravity, and smaller for a dense material like iron. A more reasonable uncertainty for typical materials is $\pm$ 20%. The assumed albedo is the largest source of uncertainty.

Integrating over the whole trail gives us the total volume of particles of $2.8 \times 10^5 \, m^3$, which corresponds to 16% of the total volume of a 120 m diameter parent body. If all the dust came from a hemispherical crater, it would have a diameter of around 80 m. Such a large crater (relative to the size of the parent body) is reasonable, as it is of similar proportions to the surprisingly large craters seen by space-craft imaging of asteroids[25]. We speculate that the survival of the parent body following such a collision strongly implies that it is a 'rubble pile'. This is also supported by the very low ejecta velocities observed, as collision experiments[12] show that these imply a low strength and high porosity target for collision speeds typical in the asteroid belt, although we note that recent computer simulations suggest that for very small asteroids even monolithic parent bodies produce low ejecta speeds[13]. An alternative explanation for the low velocity of the ejecta could be an unusually low speed collision between two asteroids with similar orbits, which is possible as the orbit of P/2010 A2 puts it within the Flora asteroid family[6], but is still highly improbable.

By assuming a density of 2500 kg m$^{-3}$ (typical value for an S-type asteroid, since the Flora family are S-types) we derive a mass of the ejecta of $3.7 \times 10^8$ kg. The power law size distribution of ejecta means that most of the volume (or mass) is contained in the largest particles closest to the parent body, so the contribution of smaller particles further along the trail (beyond the NTT/EFOSC field of view) or already lost from the trail entirely is not significant in calculating this total. The ~20% uncertainty on the conversion from $\beta$ to particle size gives a corresponding ~20% uncertainty on the total volume, but the total mass uncertainty is dominated by our choice of density for the particles. The range in possible values is ~1-6 $\times 10^8$ kg.

**Collision rates** Assuming that the parent body had an orbit similar to that of the present 120 m body, the computed parent body intrinsic average impact probability within the main belt is ~$2.9 \times 10^{-18}$ km$^{-2}$ yr$^{-1}$. The average impact velocity is ~4.8 km s$^{-1}$. These values are computed according to the best current main belt population model[19], and following the procedure recently applied to asteroid (2867) Steins[20].

Using a crater scaling law[21], it is estimated that the diameter of the impactor responsible for the formation of an 80 m crater was in the range 6-9 m, depending on the unknown strength and density of the target. We use the cohesive crater scaling law with a target density of 2000 kg m$^{-3}$ and tensile strength of $10^6 - 10^7$ dyne cm$^{-2}$ as a reasonable model for a high porosity and low strength S-type asteroid.

Therefore, the computed impact probability of the parent body with impactors having sizes of 6-9 m is about one impact every 1.1 Gyr. Considering that the main belt is estimated to be populated by some 8.6 x $10^7$ objects larger than 120 m$^{(19)}$, this implies that collisions like the observed event happen once every 12 years, approximately. This time scale is in agreement with the single discovery by the LINEAR survey.

We note that the P/2010 A2 event was discovered by LINEAR close to its detection limits, due to the faint nature of the trail. Indeed, examination of pre-discovery images by the LINEAR team revealed that the trail had been observed earlier but was missed by the automatic software that searches for new objects[26]. Therefore, as the sensitivity of the next generation of surveys will increase, it is expected that a fair number of similar discoveries will be made in the years to come. For instance, impacts in the range 3-6 m (i.e. a factor of 2 less than P/2010 A2 in size, hence a factor of 8 less in mass dust) are expected to occur every 2.5 yr on a 200 m body.

Our estimates for the P/2010 A2 event time scale depend upon the actual number of impactors in the size range 6-9 m, which is unknown since these objects are too small to be detected by present surveys. Nevertheless, extrapolation of the main belt population used in these calculations[19] to the NEO population shows that the latter fits Earth's bolides (which have diameter in the range 1-10 m)[27] within a factor of 2-3. This number can be used as an order of magnitude estimate for the uncertainty of main belt asteroids in the range 6-9 m.

The predicted dust production mass from events like the one observed for P/2010 A2 is 3-4 orders of magnitude less than the required zodiacal dust production for a steady state, and therefore in agreement with recent work suggesting that comets supply the vast majority of the zodiacal cloud[22,23]. Although beyond the scope of the present letter, we note that the total production of dust from asteroids should be obtained by integrating the contribution from all impactor and parent body sizes; accounting for the more common but smaller impacts that future surveys will find and also rarer and larger impacts.

**Supplementary References**

**Supplementary Table 1. Details of the observations.**

|  | NTT | ROSETTA | Palomar 200" | NTT |
|---|---|---|---|---|
| date/time | 2010-02-16 2UT | 2010-03-16 5-9UT | 2010-04-04 7UT | 2010-04-06 0UT |
| instrument | EFOSC2 | OSIRIS/NAC | LFC | EFOSC2 |
| r (AU) | 2.03 | 2.05 | 2.07 | 2.07 |
| Δ (AU) | 1.23 | 0.98 | 1.74 | 1.76 |
| α (deg) | 21.2 | 58.7 | 28.8 | 28.9 |
| $PA_v$ (deg) | 278.16 | 278.58 | 283.25 | 283.51 |
| ψ (deg) | 276.72 | 258.13 | 277.72 | 277.94 |
| γ (deg) | 0.49 | 10.39 | 2.44 | 2.46 |
| δ ("/hr) | 33.8 | 23.7 | 48.7 | 49.2 |
| $t_{exp}$ (s) | 600 | 870 | 360 | 300 |
| $N_{exp}$ | 5 | 16 | 2 | 3 |
| filter | R | clear | R | R |
| pixel ("/km) | 0.24/214 | 3.8/2700 | 0.36/457 | 0.24/306 |
| $PA_{mean}$ (deg) | 278.3 ± 0.1 | 320.7 ± 0.5 | 286.4 ± 0.1 | 285.4 ± 0.1 |

Note. The date and time of each observation are summarized together with the distance from the Sun (r) and from the observer (Δ), and the phase angle (α) at the time of the observations. $PA_v$ is the position angle of the heliocentric velocity vector (i.e. orbit) of the object projected in the sky measured counter-clockwise North over East, ψ indicates the anti-sunward direction and γ is the angle between the observer and the target orbital plane. δ is the total rate of motion relative to the stars in arc-seconds per hour. From Earth the motion was mostly towards the East, from Rosetta it was towards the South-East. The exposure time, the number of exposures and the filter used for the observations are summarized, and the pixel scale given in both arc-seconds and km (projected on sky at the distance of P/2010 A2). The last row contains the position angle of the trail as measured in our frames.

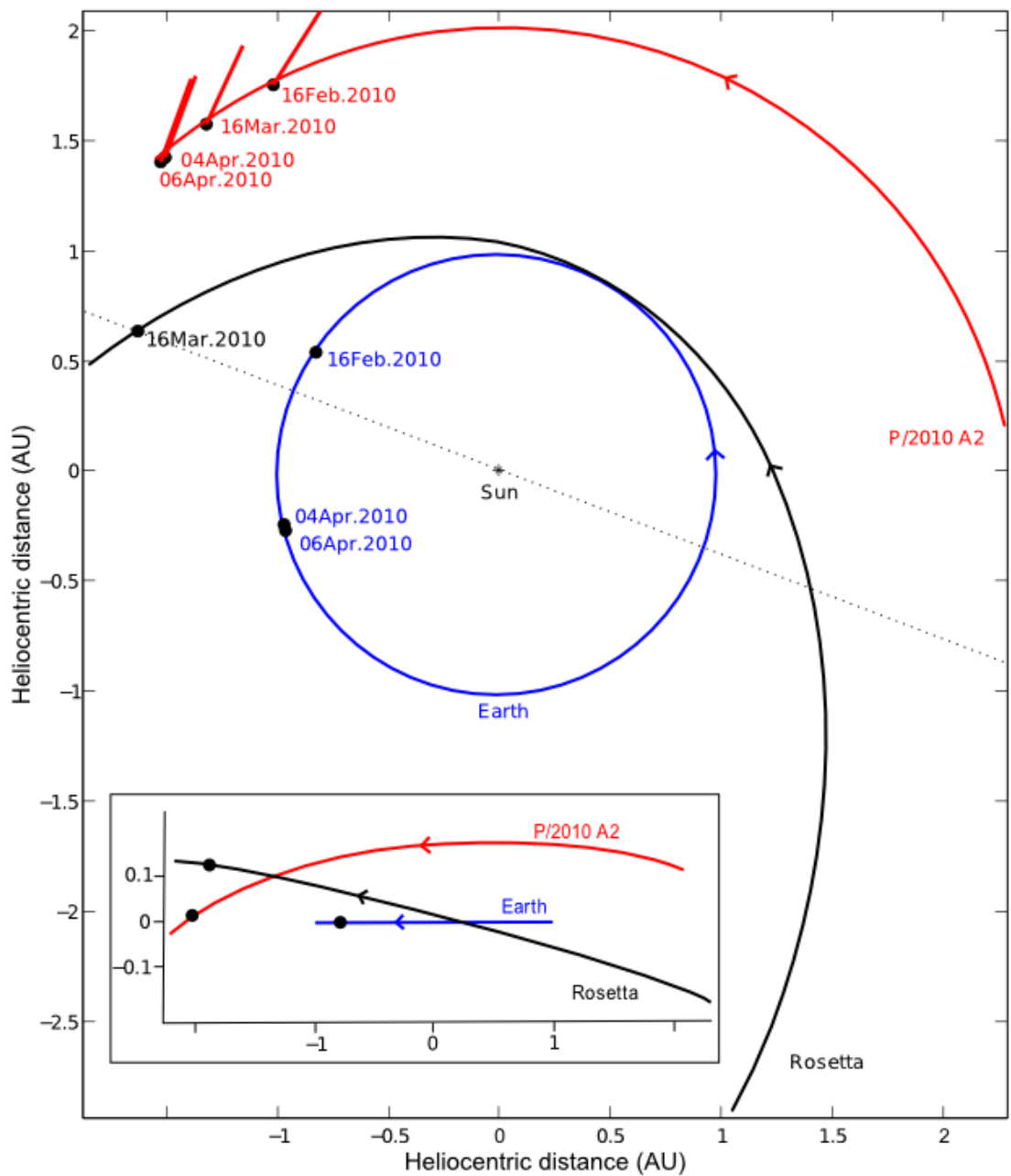

**Supplementary Figure 1**. The orbits of the Earth, Rosetta and P/2010 A2. Dots represent the positions at the time of observations. Thick lines indicate the direction of the dust trail in space at each epoch (length not to scale). The inset shows a cross section (along the dotted line) showing the orbital planes of P/2010 A2 and Rosetta relative to the ecliptic (scales also in AU), with the points showing the positions at the time of the Rosetta observations.

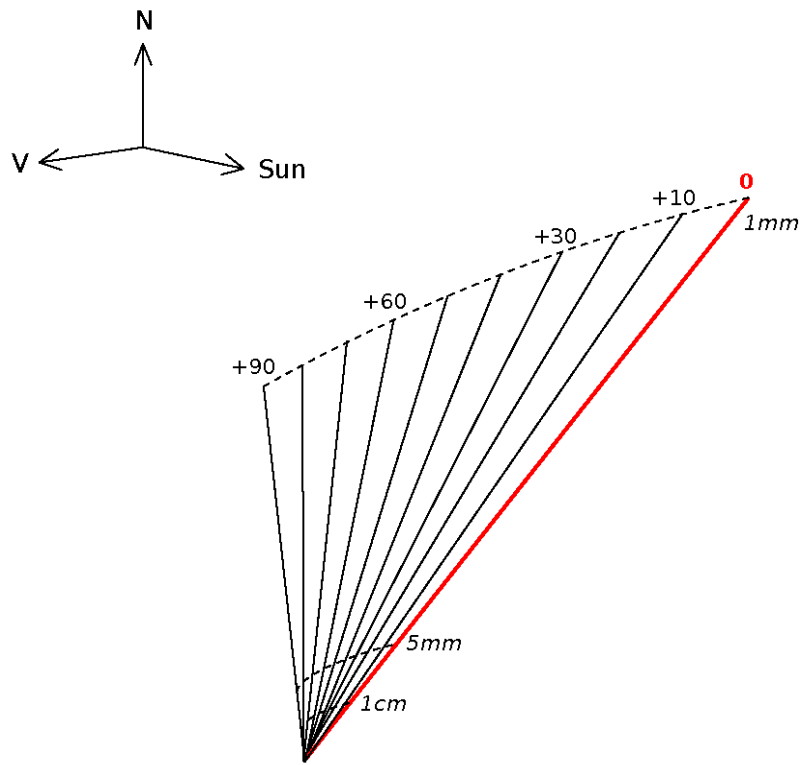

**Supplementary Figure 2**. Finson-Probstein model showing a simulated image for the OSIRIS observing geometry. The synchrones are labelled with the time in days since the start of activity (10[th] February 2009), while the syndynes are labelled with the diameter of particles corresponding to the *β* value at that distance. The compass in the top-left shows the orientation of the image (North up, East left, as viewed on sky), the direction V of the velocity vector (orbital motion) of P/2010 A2 and the direction to the Sun. This model has large particles (mm – cm) and ongoing activity over an extended period (a comet model). The simulated OSIRIS image shows that such activity would produce a fan shaped tail, which can be ruled out by the real image. From an Earth-based geometry, the trail would appear as a straight line in this model, matching the observations. This is the case for all models, so the simulated Earth based views are not shown as they cannot rule out scenarios.

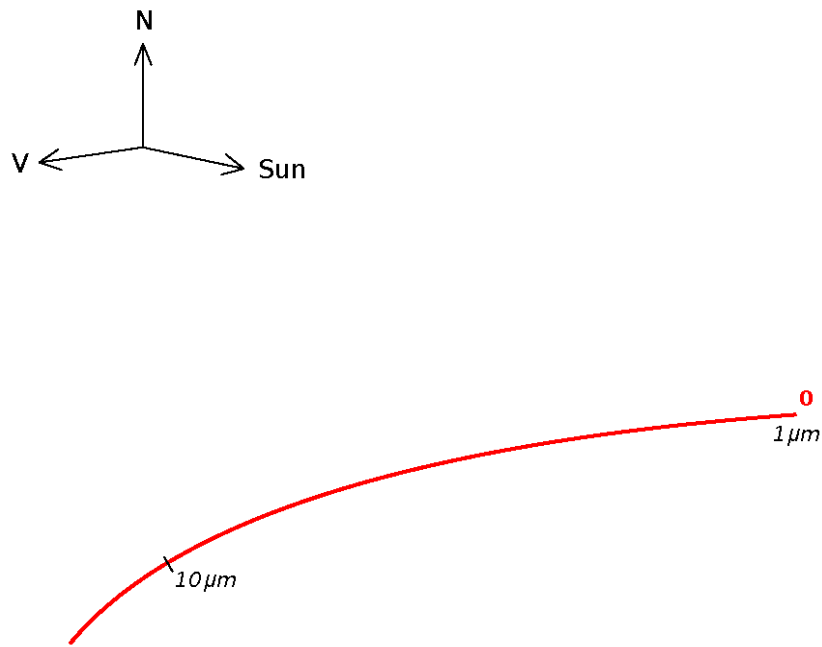

**Supplementary Figure 3**. Finson-Probstein model showing a simulated image for the OSIRIS observing geometry. This model has small particles (micron – mm) and a burst of activity over a short period (a collision model). It produces a narrow arc with a strong curvature rather than the straight synchrones seen in the large particle model. This is also clearly different from the observed trail.

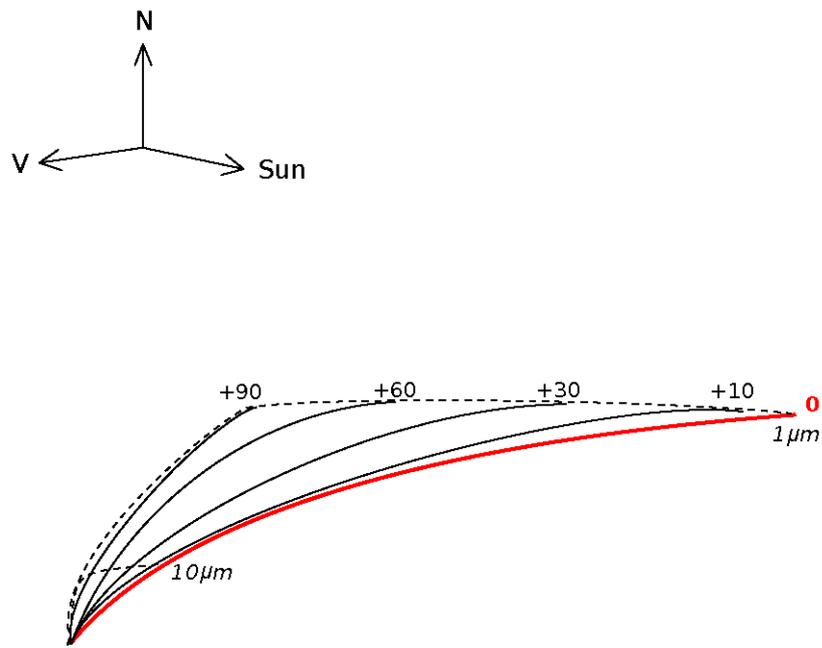

**Supplementary Figure 4**. Finson-Probstein model showing a simulated image for the OSIRIS observing geometry. This model has small particles (micron – mm) and ongoing activity over an extended period (a comet model). This produces a strongly curved fan of material, and is ruled out by the observations.

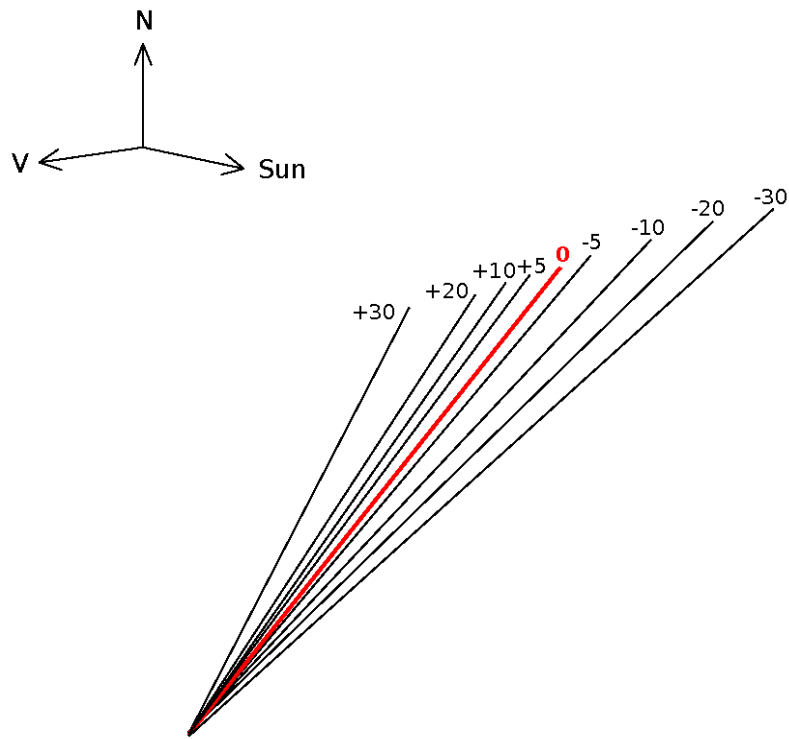

**Supplementary Figure 5**. Finson-Probstein model showing simulated images for the OSIRIS observing geometry. This set of models have large particles (mm – cm) and a burst of activity over a short period (collision models). We plot synchrones based on collisions on a variety of dates (times are given in days relative to 0 UT on 10 February 2009) to demonstrate the accuracy of the date determination. Based on the accuracy of the PA measurement in the OSIRIS image, we can constrain the date of the collision to within +/- 5 days.

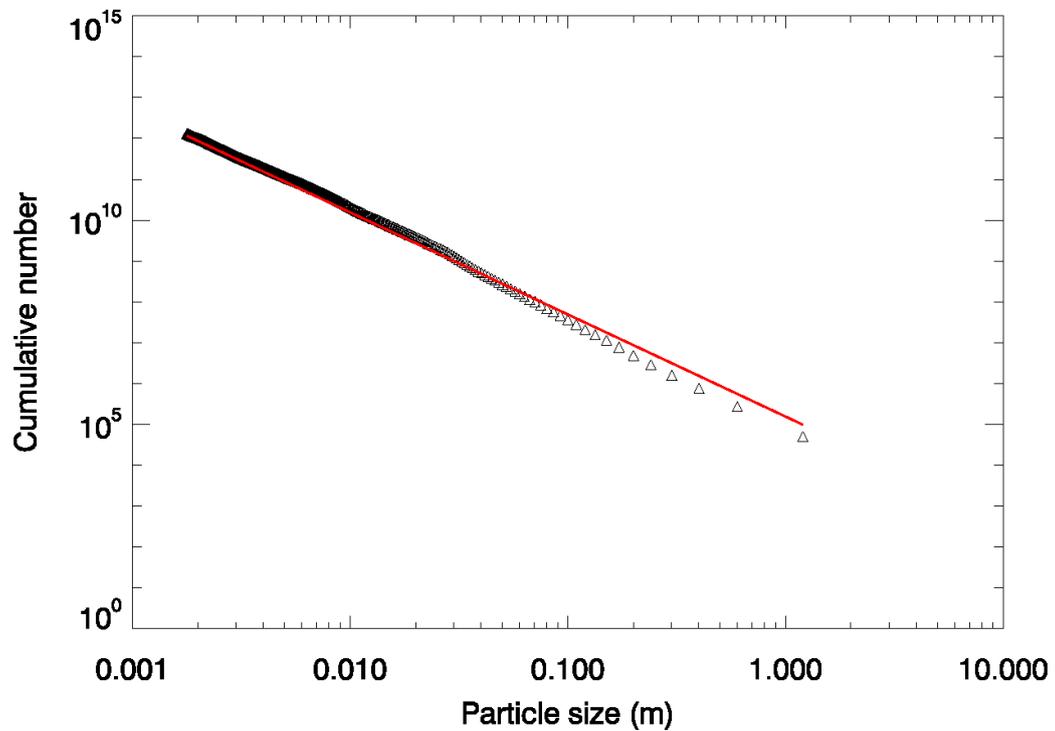

**Supplementary Figure 6**. Cumulative size distribution of ejecta particles. The number of particles larger than a given radius is shown. The distribution has a slope near to $q = 2.5$ as expected for a population of collisional remnants (shown by the red line). The number of particles was calculated from the flux profile in the NTT image and the size of particles at each distance along the trail from the Finson-Probstein model.